\documentclass[aps,pre,reprint,amsmath,amssymb,showpacs]{revtex4-1}
\usepackage{graphicx}

\begin{document}


\title{Brownian dynamics of a self-propelled particle in shear flow}


\author{Borge ten Hagen}
\email[]{bhagen@thphy.uni-duesseldorf.de}
\affiliation{Institut f\"ur Theoretische Physik II: Weiche Materie, Heinrich-Heine-Universit\"at D\"usseldorf, Universit\"atsstr.~1, D-40225 D\"usseldorf, Germany}
\author{Raphael Wittkowski}
\affiliation{Institut f\"ur Theoretische Physik II: Weiche Materie, Heinrich-Heine-Universit\"at D\"usseldorf, Universit\"atsstr.~1, D-40225 D\"usseldorf, Germany}

\author{Hartmut L\"owen}
\affiliation{Institut f\"ur Theoretische Physik II: Weiche Materie, Heinrich-Heine-Universit\"at D\"usseldorf, Universit\"atsstr.~1, D-40225 D\"usseldorf, Germany}


\date{\today}

\begin{abstract}
Brownian dynamics of a self-propelled
 particle in linear shear flow
is studied analytically by solving the Langevin equation and in simulation. The particle has a constant propagation 
speed along a fluctuating orientation and is additionally subjected to
a constant torque. In two spatial dimensions, the mean trajectory  and the 
mean square displacement (MSD) are calculated as 
functions of time $t$ analytically. In general, the mean trajectories are cycloids that are modified by finite temperature effects. With regard to the MSD different  regimes are identified
where the MSD  scales with $t^\nu$ with  $\nu = 0,1,2,3,4$. 
In particular,  an accelerated
($\nu =4$) motion emerges if the particle
is self-propelled along the gradient direction of the shear flow. 
\end{abstract}

\pacs{82.70.Dd, 05.40.Jc}

\maketitle


\section{\label{Einleitung}Introduction}
Mesoscopic colloidal particles which perform Brownian dynamics in a viscous solvent
exhibit intriguing non-equilibrium behavior if they are exposed to shear flow. In fact, shear flow drastically affects their viscoelastic response \cite{Kalman:09,Siebenbuerger:09,Prasad:09,Besseling:09}, 
criticality and phase behavior \cite{Dhont:98,Lettinga:04,Loewen:01,Blaak:04}, 
and other collective effects with \cite{Gompper:09,Iwashita:09} and 
without \cite{Krueger:09,Krueger:10} many-body hydrodynamic interactions.
As an example, the mean square displacement of a Brownian particle in shear flow contains 
non-vanishing cross-correlations \cite{vandeVen_book,Ziehl:09} 
and shows several dynamical regimes when confined 
 to an additional parabolic potential \cite{Holzer:10}.

Recently, \textit{active} particles which are self-propelled by their 
own intrinsic motor~\cite{Ramaswamy,Lauga_review:09,Popescu:09} have been studied.
Even the single-particle dynamics is a non-equilibrium situation since the particles dissipate energy.
Most of the recent studies of self-propelled particles were performed in a quiescent 
solvent (see, e.g., Refs.\ \cite{Howse:07,Teeffelen_PRE,Cond_Matt,Dunkel:09,Lobaskin:08,Peruani:07,Teeffelen:09}). 
But also studies  focused on self-propelled particles in an imposed shear field \cite{Muhuri:07,Cates:08,Pahlavan:11,Koch:11} gain more and more interest. Apart from its fundamental importance,
the interplay between hydrodynamic flow and active particle dynamics is also relevant for ecology,
having for instance the motion of bacteria in oceans, lakes, and rivers in mind \cite{Mitchell:06}. Besides particles with a fixed shape, also swimmers which change their shape during propulsion were considered~\cite{Ohta:09} or a velocity field due to surface deformations was prescribed on the surface of the particle~\cite{Downton:09,Goetze:10}.

In this paper, we study the two-dimensional Brownian dynamics~\cite{Han:06} of a single self-propelled 
spherical particle in linear shear flow based on the Langevin equations.
The particle has a constant propagation speed which fluctuates in its direction and is exposed to
a linear shear flow and a constant torque. 
Analytical results as 
functions of time $t$ are presented for  the mean trajectory and the
mean square displacement (MSD).  At zero temperature the mean trajectories are cycloids, which are modified if finite temperature effects are included. For the MSD different  regimes are identified
where the MSD  scales with $t^\nu$ with  an exponent $\nu = 0,1,2,3,4$.
While $\nu=3$ is the most general case for long times,
the other exponents are realized in the following special cases:
$\nu=0$ in the absence of shear flow \textit{and} fluctuations,
$\nu=1$ in the absence of shear flow,
$\nu=2$ in the absence of shear flow,  torque,  \textit{and} fluctuations,
$\nu=4$ in the absence of fluctuations \textit{and} for a special nonzero torque 
canceling the shear rotation. Live unicellular motile microalgae (\textit{Chlamydomonas Reinhardtii}) that maintain their direction in shear flow~\cite{Rafai:10} provide an experimental realization of particles that resist the flow rotation for most of the time. 
The accelerated motion
with the  exponent $\nu =4$  is found  if a particle 
is self-propelled along the gradient direction of the shear flow.
As a general result, the motion of self-propelled particles is greatly amplified by shear
and, for special initial conditions, the constant acceleration can hugely enhance the mobility
of individual particles.  Our results can in principle be verified in experiments of self-propelled colloidal particles
in shear flow \cite{Baraban:08}.  

The model considered in this paper has been solved before in two special limits.
In the absence of shear, the solution was presented in Ref.\ \cite{Teeffelen_PRE}, while
 in the absence of self-propulsion, the traditional Brownian motion in shear flow
is obtained \cite{Cerda:83,Ziehl:09}. It is important, however, that in none 
of these limits the exponent 4 emerges which is thus arising from a combination of self-propulsion 
and shear.

In addition to the analytical analysis of mean trajectory and MSD, in this paper, we also focus on the complete probability distribution function for the displacement of a self-propelled particle in shear flow. As calculated in simulation, it exhibits a transient double-peak structure, which is due to the self-propulsion and  distorted by the shear flow. 

This paper is organized as follows: In Sec.\ \ref{model} we introduce our model of a self-propelled particle in shear flow. The analytical solutions are presented for the special case of zero temperature in Sec.\ \ref{Tzero} and for the general case of finite temperature in Sec.\ \ref{Tfinite}. Section \ref{simulation} contains some further results obtained by simulation and, finally, a conclusion is given in Sec.\ \ref{conclusion}.

\section{\label{model}The model}

 \begin{figure}
 \centering
\includegraphics[width=\columnwidth]{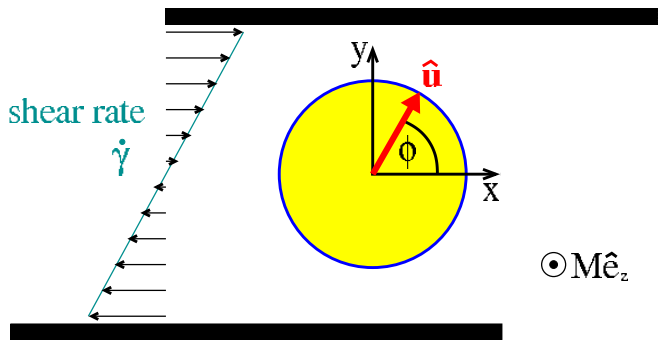}%
\caption{\label{fig:Modell}(Color online) Sketch of the self-propelled particle in Couette flow with shear rate $\dot{\gamma}$. The system is  considered in the two-dimensional $(x,y)$ plane. A systematic torque $\mathbf M= M \mathbf {\hat{ e}_z} $  and the direction $\mathbf {\hat{ u}}$  of the effective force $\mathbf F$ due to the self-propulsion are indicated.}
\end{figure}
In our model, we consider the completely overdamped two-dimensional Brownian motion of a spherical self-propelled particle in Couette flow (see Fig.\ \ref{fig:Modell}). The distance between the boundary plates and the particle is assumed to be large enough so that hydrodynamic interactions can be neglected.  The mechanism of propulsion is taken into account by means of an effective internal force $\mathbf F= F \mathbf {\hat{ u}} = F \left( \cos (\phi), \sin (\phi) \right) $, which enters the Langevin equations. The centre-of-mass position vector $\mathbf{r}= (x,y)$    and the angle $\phi$ between the unit vector $\mathbf {\hat{ e}_x}$ that is co-directional with the Cartesian $x$ axis and the orientational unit vector $\mathbf {\hat{ u}}=(\cos(\phi),\sin(\phi))$ are used to describe the two-dimensional motion of the self-propelled particle. The effect of the shear flow enters the equations via the shear rate $ \dot{\gamma}$. Considering a systematic torque $M$ as indicated in Fig.\ \ref{fig:Modell}, a Gaussian white noise random force $\mathbf f = \left( f_x, f_y\right)$ and a Gaussian white noise random torque $g$ leads to the following set of Langevin equations:  
\begin{align}
\label{pxGleichung}
\frac{\mathrm{d}x}{\mathrm{d}t} &= \dot{\gamma} y + \beta D_t \left(F \cos (\phi) +f_x\right)\,,\\
\label{pyGleichung}
\frac{\mathrm{d}y}{\mathrm{d}t} &= \beta D_t \left( F \sin (\phi) +f_y \right)\,,\\
\label{pphiGleichung}
\frac{\mathrm{d}\phi}{\mathrm{d}t} &= -\frac{ \dot{\gamma}}{2} +\beta D_r \left( g + M \right) \,.
\end{align}
Here, $\beta=1/(k_\mathrm{B}T)$ is the inverse effective  thermal energy and $D_t$ and $D_r$ are the translational and
rotational short-time diffusion constants satisfying the ratio  $D_t/D_r = 4 R^2/3$ for spherical particles with radius $R$. With $\langle \,\cdot\, \rangle$ denoting the noise average, the Gaussian white noise quantities are characterized by $\langle f_i(t)\rangle =0$, $\langle f_i(t)
f_j(t')\rangle =2\delta_{ij}\delta(t-t')/(\beta^2 D_t)$, $\langle g(t)\rangle =0$, and $\langle g(t)g(t')\rangle =2\delta(t-t')/(\beta^2 D_r)$, where $i,j \in \{x,y\}$ and $\delta_{ij}$  is the Kronecker delta symbol. Here, we assume out of equilibrium noise to be close to the equilibrium one.

The subsequent analysis in this paper shall be carried out in a dimensionless form. For that purpose, we introduce the following reduced parameters: $\tilde{x}= x/R$, $\tilde{y}= y/R$, $\tau = D_r t$, $\tilde{f}_i = f_i/F$, $\tilde{g}_i = g/M$, $\alpha= 4 \beta R F / 3$, $\mu =\beta M$, and $\xi= \mathrm{Pe_r}/2 =\dot{\gamma}/(2D_r)$, where $\mathrm{Pe_r}=\dot{\gamma}/D_r$ is the rotational P\'{e}clet number.
With these parameters the new set of dimensionless Langevin equations can be written as 
\begin{align}
\label{xGleichung}
\frac{\mathrm{d}\tilde{x}}{\mathrm{d}\tau} &= 2 \xi \tilde{y} + \alpha \bigl( \cos (\phi) +\tilde{f}_x\bigr)\,,\\
\label{yGleichung}
\frac{\mathrm{d}\tilde{y}}{\mathrm{d}\tau} &= \alpha \bigl( \sin (\phi) +\tilde{f}_y \bigr)\,,\\
\label{phiGleichung}
\frac{\mathrm{d}\phi}{\mathrm{d}\tau} &= -\xi +\mu\left( 1 + \tilde{g}\right) \,.
\end{align}

\section{\label{Tzero}The case of zero temperature}
At zero temperature, as there is no thermal motion, the noise terms in the system of dimensionless Langevin equations~(\ref{xGleichung})-(\ref{phiGleichung}) can be neglected and an analytical solution for the trajectory can be given. One has to distinguish between the cases $\mu=\xi$ and $\mu \neq \xi$.

For $\mu \neq \xi$ the reduced net torque $\omega$ given by $\omega=\mu-\xi$ acts on the particle. This leads to the analytical results 
\begin{align}
\label{xsungleichr}
\Delta \tilde{x} &=  2 \xi \left( \tilde{y}_0 + \frac{\alpha \cos (\phi_0)}{\omega} \right) \tau \notag \\
&\quad  + \frac{\alpha(3 \xi - \mu)}{\omega^2} \left[\sin (\phi_0) -  \sin \left( \phi_0 + \omega \tau \right) \right]   \,,  \\
\label{ysungleichr}
\Delta \tilde{y} &= \frac{\alpha}{\omega} \left[ \cos (\phi_0) - \cos \left( \phi_0 + \omega \tau \right) \right]\,,
\end{align}
where $\Delta \tilde{x} =  \tilde{x}(\tau)  -  \tilde{x}_0 $ and $\Delta \tilde{y} = \tilde{y}(\tau)  -  \tilde{y}_0$ with the initial reduced centre-of-mass position vector $\mathbf{\tilde{r}_0} \equiv \mathbf{\tilde{r}}(t=0) = (\tilde{x}_0,\tilde{y}_0)$  and the initial orientation angle $\phi_0 \equiv \phi(t=0)$ of the self-propelled particle.
Equations~(\ref{xsungleichr}) and (\ref{ysungleichr}) describe \textit{cycloids} as can also be seen from the corresponding graph (dashed line) in Fig.\ \ref{fig:Tfinite1}.

For the special case $\mu=\xi$ we get
\begin{align}
\label{xsgleichr}
\Delta \tilde{x} &=   \xi \alpha \sin (\phi_0) \tau^2  +  \left( 2 \xi \tilde{y}_0  +   \alpha \cos (\phi_0) \right) \tau \,, \quad \\
\label{ysgleichr}
\Delta \tilde{y} &=  \alpha  \sin (\phi_0) \tau \,,
\end{align}
Physically, the condition $\mu=\xi$ implies that the rotation of the self-propelled particle due to the shear flow is exactly compensated by the additional external or internal torque represented by $\mu$. Thus, the orientation of the particle remains constant all the time as there is no random torque in the case of zero temperature, either.

\section{\label{Tfinite}Generalization to the case of finite temperature}
\subsection{Mean trajectory}
 \begin{figure}
 \centering
\includegraphics[width=\columnwidth]{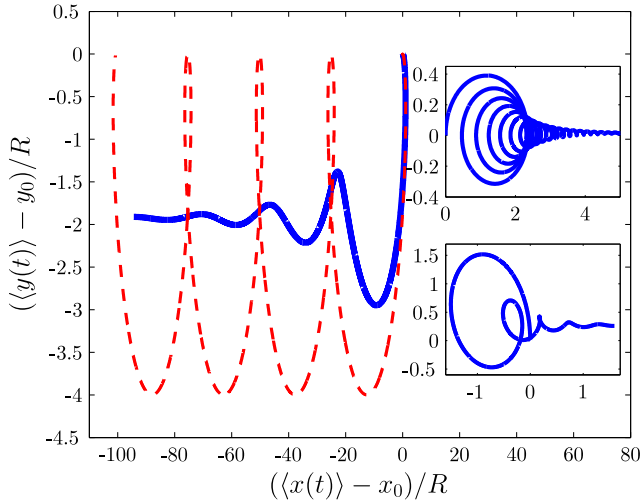}%
\caption{\label{fig:Tfinite1}(Color online) Mean trajectory of a self-propelled particle in shear flow at finite temperature (solid line in the main figure) and the corresponding trajectory for $T=0$ (dashed line).  The damping effect of the finite temperature is clearly visible.  The parameters are $\alpha = 10$ and $\phi_0 = 0$ for all plots, while the normalized shear rate is $\xi=5$ in the main figure and $\xi=25$ in the upper inset. In the lower inset besides $\xi=1$, an additional torque determined by $\mu=7$ is considered as well.   }
\end{figure}
For finite temperature $T>0$ the Brownian motion due to collisions of the self-propelled particle with solvent particles has to be taken into account. 
The mean trajectory of the self-propelled particle can then be written in the form
\begin{align}
\left\langle \Delta \tilde{x}\right\rangle & =  2 \xi \! \left( \! \frac{\alpha \phi_0^+}{1+\omega^2} + y_0  \! \right) \!\tau + \frac{\alpha \phi_0^\pm}{1+\omega^2}  - \frac{\alpha \Delta \phi^\pm}{1+\omega^2}  e^{-\tau} \!, 
\label{eq:1xScherung} \\
\left\langle \Delta \tilde{y}\right\rangle & = \frac{\alpha \phi_0^+}{1+\omega^2} - \frac{\alpha \Delta \phi^+}{1+\omega^2} e^{-\tau}  \,.
\label{eq:1yScherung}
\end{align}
Here,  the parameters $\phi_0^+$ and $\phi_0^\pm$ containing the dependence on the initial angle $\phi_0$ were used. They are given by 
\begin{align}
\label{phi+}
\phi_0^+  & = \sin (\phi_0) + \omega \cos (\phi_0) \,,\\
\label{phi-}
\phi_0^- & = \cos (\phi_0) - \omega \sin (\phi_0) \,,\\
\label{phi+-}
\phi_0^\pm &  = \phi_0^- - \frac{2\xi}{1+\omega^2} \left( \phi_0^+ + \omega \phi_0^-\right) .
\end{align}
The parameters $\Delta \phi^+$, $\Delta \phi^-$, and $\Delta \phi^\pm$ are defined in exactly the same way, but with $\Delta \phi$ instead of  $\phi_0$  in Eqs.~(\ref{phi+})-(\ref{phi+-}). The analytical results  for the mean trajectory are visualized by the solid lines in Fig.\ \ref{fig:Tfinite1} and its insets. The main part of this figure illustrates basically two effects due to the finite temperature. Comparing the dashed line corresponding to $T=0$ with the solid line shows that the finite temperature leads first to a damping in the amplitude and second to a reduced frequency of the oscillations. As presented in the lower inset, an additional torque leads to quite complicated mean trajectories.

\subsection{Mean square displacement}
 \begin{figure}
 \centering
\includegraphics[width=\columnwidth]{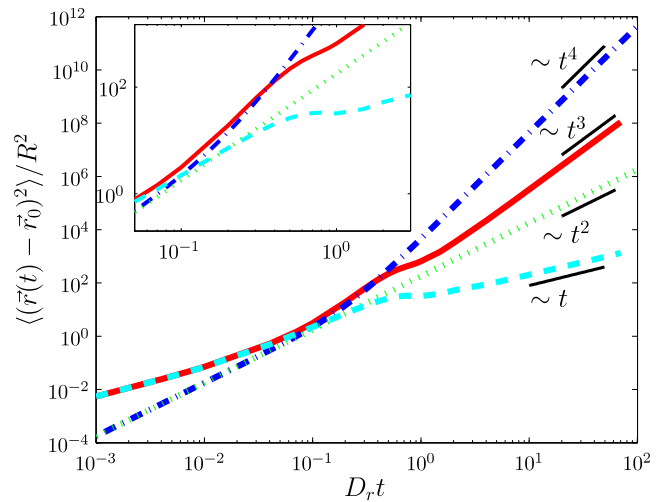}%
\caption{\label{fig:Tfinite2}(Color online) Mean square displacement of a self-propelled particle in linear shear flow. For all plots the strength of the self-propulsion is determined by $\alpha = 10$ and the initial angle is $\phi_0 = \pi/2$. While $T_1=0$ for the dotted and the dashed-dotted lines, the solid and the dashed lines refer to situations with finite temperature $T_2>0$. A nonzero shear flow ($\xi=5$) is considered for the solid and the dashed-dotted lines. An additional torque ($\mu = 5$) is taken into account for the dashed and the dashed-dotted lines. Depending on the temperature and the existence of  shear flow and additional torque, different exponents of $t$ can be identified in the long-time behavior. In the inset a close-up of the intermediate regime is shown.}
\end{figure}
 Next, the MSD is discussed. The $x$ and $y$ components can be considered separately. The full analytical solution, which is available, contains an enormous number of terms and is far too long to be given here explicitly. Therefore, we  limit ourselves to presenting the integrals and correlation functions that have to be calculated:
 \begin{align}
& \left \langle (\Delta \tilde{x})^2 \right\rangle  = \frac{32}{9} \xi^2 \tau^3 + 4 \xi^2 {\tilde{y}_0}^2 \tau^2  + \frac{8}{3} \tau  \notag \\
& \quad +  4 \alpha^2 \xi^2 \!\! \int_0^{\tau} \!\! \!\! \mathrm d\tau_1 \!\! \int_0^{\tau_1} \!\!\!\!\!  \mathrm d\tau_2 \!\! \int_0^{\tau} \!\!\!\!  \mathrm d\tau_3 \!\! \int_0^{\tau_3}  \!\!\!\!\!   \mathrm d\tau_4  \left \langle \sin  \left(\phi\left(\tau_2\right)\right) \sin \left(\phi\left(\tau_4\right)\right) \right\rangle \notag \\
& \quad  + \alpha^2 \!\! \int_0^{\tau} \!\! \!\! \mathrm d\tau_1 \!\! \int_0^{\tau} \!\!\!\!  \mathrm d\tau_2 \left \langle \cos \left(\phi\left(\tau_1\right)\right) \cos \left(\phi\left(\tau_2\right)\right) \right\rangle
 \notag \\
& \quad +  8 \alpha \xi^2 \tilde{y_0} \tau \!\! \int_0^{\tau} \!\! \!\! \mathrm d\tau_1 \!\! \int_0^{\tau_1} \!\!\!\!  \mathrm d\tau_2 \left \langle \sin \left(\phi\left(\tau_2\right)\right) \right\rangle \notag \\
& \quad + 4 \alpha \xi \tilde{y_0} \tau  \!\! \int_0^\tau  \! \! \!\! \mathrm d\tau_1 \left \langle \cos \left(\phi\left(\tau_1\right)\right) \right\rangle  \notag \\
& \quad
+ 4 \alpha^2 \xi \!\! \int_0^{\tau} \!\! \!\! \mathrm d\tau_1 \!\! \int_0^{\tau_1} \!\!\!\!\!  \mathrm d\tau_2 \!\! \int_0^{\tau} \!\!\!\!  \mathrm d\tau_3  \left \langle \sin \left(\phi\left(\tau_2\right)\right) \cos \left(\phi\left(\tau_3\right)\right) \right\rangle .
\label{MSD}
\end{align}
Carrying out the calculations  reveals  a rich variety of regimes as visualized in Fig.\ \ref{fig:Tfinite2}.    This is particularly remarkably because the underlying model situation is quite simple. In the following, situations that lead to the different exponents $\nu = 0,1,2,3,4$ shall be covered in more detail. 

The case $\nu = 0$ is realized for $T=0$ in the absence of shear flow. The corresponding motion of the self-propelled particle describes a closed circle due to a nonzero torque. The exponent $\nu=1$ is the most general case without shear flow. Details of this model for a self-propelled particle were presented in Ref.\ \cite{tenHagen:11}. The ballistic case $\nu =2$ is more or less trivial and is realized for a zero torque at $T=0$ in the absence of shear flow. The particle simply moves on a straight line. The most comprehensive situation revealing the most interesting physics consists of systems with shear flow and torque at finite temperature $T$ and in general leads to $\nu =3$. In this case, basically three regimes can be identified. For short times, the MSD is linear in time as the simple diffusive motion is dominant. As soon as the self-propulsion becomes significant, a crossover regime is observed, where contributions with $t^2$, $t^3$, and $t^4$ are relevant. A more detailed discussion of this regime is given below. Finally, at the time scale $1/D_r$, the intermediate regime is terminated by the long-time $t^3$ law.   While the exponent $\nu = 3$ also occurs in the Taylor diffusion of passive particles~\cite{Cerda:83}, the exponent $\nu = 4$, which is found for the situation with a torque exactly canceling the shear rotation also in the long-time behavior (see dashed-dotted line in Fig.\ \ref{fig:Tfinite2}), is characteristic and only realized for active particles. 

In the inset of Fig.\ \ref{fig:Tfinite2} a close-up of the crossover regime is shown.   While the $t^2$ and $t^3$ terms in this transient regime also occur for self-propelled particles in  a quiescent solvent~\cite{tenHagen:11}, the $t^4$ contribution is only found for situations with shear flow. If the initial orientation of the particle is parallel to the $y$ direction ($\phi_0 = \pi/2$) and the shear rotation is exactly compensated by the additional torque ($\mu = \xi$), the motion is directed in the $y$ direction for a significant time. Thus, an accelerated motion with the exponent $\nu =4$ arises. Unlike the case of zero temperature, for nonzero noise the $t^4$ contribution is restricted to a transient regime, which is terminated by the general $t^3$ law. 
Although the condition $\mu = \xi$ seems to be very special at first sight, an example of particles that resist the flow rotation for most of the time is given by live unicellular motile microalgae (\textit{Chlamydomonas Reinhardtii}) that maintain their direction in shear flow~\cite{Rafai:10}. Thus, the condition $\mu = \xi$ is realized in experiment. 

The exact expression for the long-time behavior in the most general case ($\nu = 3$) is given by the first term in Eq.~(\ref{MSD}) for passive particles. Depending on the strength of the self-propulsion and the effective torque in the shear flow, for self-propelled particles a second contribution becomes relevant:   
\begin{equation}
\lim_{t\rightarrow\infty}\langle(\Delta \tilde{x})^2\rangle = \frac{32}{9} \xi^2 \left[1 + \frac{3}{8}\frac{\alpha^2}{\left(1+ \omega^2 \right) }\right] \tau^3 \,.
\label{thoch3}
\end{equation}
Going back to physical quantities, for the case of passive particles the leading term can be written as $(2/3) \dot{\gamma}^2 D_t t^3$. Equation~(\ref{thoch3}) verifies that the corresponding term for self-propelled particles is obtained by replacing the translational diffusion constant $D_t$ for passive particles with the substantially enhanced long-time translational diffusion constant~\cite{tenHagen:11,Howse:07} for self-propelled particles with net torque $\omega$.

\section{\label{simulation}Computer simulation}
 \begin{figure}
 \centering
\includegraphics[width=\columnwidth]{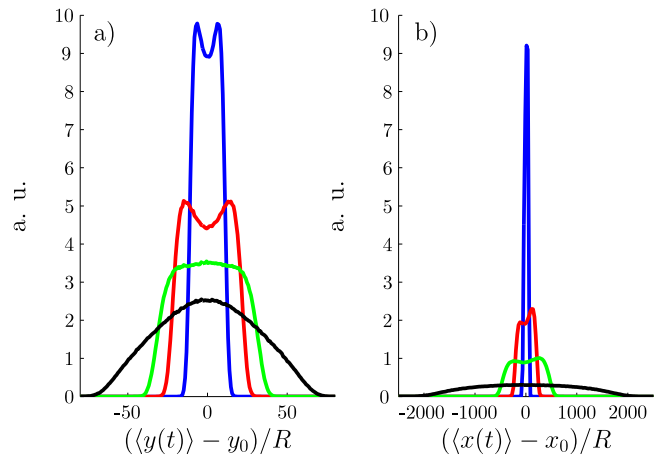}%
\caption{\label{fig:tVerteilung}(Color online) Time evolution of the probability distribution function of a self-propelled particle. In plot (a) the distribution in the $y$ direction is shown while in plot (b) the additional effect due to the shear flow in the $x$ direction is illustrated. The different lines in blue, red, green, and black correspond  to $\tau=1$, $\tau=2$, $\tau=3$, and $\tau=6$, respectively. The initial orientation of the particle points in the positive $x$ direction.}
\end{figure}
In addition to the analytical considerations presented so far, some further  information can be gained by numerical computer simulation. While the different moments of the probability distribution function for the displacement of a self-propelled particle in shear flow -- in particular the mean position and the MSD -- can be calculated analytically, this is not the case for the probability distribution function itself. Thus, we make use of a Brownian dynamics simulation here. Based on the Milstein method, a very good approximation for the probability distribution function is obtained by averaging over a huge number of particle trajectories~\cite{KloedenP2006}. Some results for the case that the rotational motion due to shear is exactly compensated by the additional torque are presented in Fig.\ \ref{fig:tVerteilung}. The left plot (a) shows the distribution in the $y$ direction, which is qualitatively different from the particle distribution in the $x$ direction [see Fig.\ \ref{fig:tVerteilung} (b)] due to the fact that the consequences of the shear flow become obvious in the latter case. The  double-peak structure, which is symmetric in the left-hand plot, can be explained as follows: Given that the active particle is oriented in the $x$ direction at the beginning, no motion in the $y$ direction due to the self-propulsion occurs at very short times. The important aspect is that at finite temperature the particle does not only undergo translational, but also rotational Brownian motion. This results in a randomly changing particle orientation and thus gives rise to a varying $y$ component of the propulsive motion. The rotational motion of the particle at very short times due to collisions with the solvent molecules determines whether the self-propulsion drives the particle in the positive or negative $y$ direction. As seen in the left plot in Fig.\ \ref{fig:tVerteilung}, the resulting double-peak is clearly observed at times $D_rt<3$. For larger times, this detailed structure is more and more washed out and, finally, for very large times a Gaussian distribution function is established. Graphically speaking, this is the case when the initial change of the particle orientation becomes irrelevant due to the continuing rotational Brownian motion. As already discussed in Sec.\ \ref{Tfinite}, in this long-time regime the motion becomes diffusive again with a modified diffusion constant due to the self-propulsion of the particle. 

On top of the previous considerations, the effect of the shear flow becomes relevant with regard to the case shown in Fig.\ \ref{fig:tVerteilung} (b). The form of the probability distribution function is not symmetric any more because the initial orientation of the particle points in the positive $x$ direction and the shear flow induces an additional component of translational motion in positive $x$ direction for positive $y$ values and in negative $x$ direction for negative $y$ values of the current particle position. This explains the modified and asymmetrical form of the particle distribution function in $x$ direction.

\section{\label{conclusion}Conclusion}

In conclusion, within an analytical solution of a two-dimensional model we have studied the mean trajectory and the 
scaling behavior of the mean square displacement (MSD) as a function of time if a self-propelled 
particle is exposed to linear shear flow. For long times, the MSD scales with $t^3$ as a result of a combined
diffusion and convection. If the self-progagation is initially oriented in the gradient direction of the 
shear flow, there is a constant-acceleration behavior where the MSD scales with $t^4$, which finally crosses over
to the $t^3$ law where the  crossover time corresponds to the inverse rotational diffusion constant.
The scaling behavior of the MSD and the transient double-peak structure of the probability distribution function for the displacement of the particle as calculated in simulation can be tested in experiments \cite{Baraban:08,Leptos:09}.

For future work, it is interesting to generalize the present solution 
to the case of oscillatory shear \cite{Heyes:94,Dhont:98,Christopoulou:09} and to three spatial dimensions
\cite{tenHagen:11}.
Further future work should focus on the behavior of an ensemble of self-propelled particles which
are coupled by both direct interactions and by hydrodynamic interactions mediated by the solvent~\cite{Baer,Wensink:08,Gompper:09}. At finite densities, deviations from a Gaussian behavior~\cite{Leptos:09} in the motion of particles are expected due to the power-law decay of hydrodynamic interactions~\cite{Dunkel:10}. The swimmer-tracer scattering~\cite{Dunkel:10} of a self-propelled particle in shear flow might be interesting as well. 
An external imposed shear flow is expected to change the viscoelastic behavior drastically \cite{Muhuri:07,Cates:08}
and there is an intricate  coupling between the intrinsic flow made by the motion of the active particles 
and the externally imposed shear flow.

\begin{acknowledgments}
We thank Adam Wysocki for helpful discussions. This work was supported by the DFG (SFB TR6 - C3).
\end{acknowledgments}

\bibliography{journals,literature}

\end{document}